\documentclass[twocolumn]{aastex62}
\usepackage[varg]{txfonts}
\usepackage{graphicx}
\usepackage{color}
\usepackage{eqnarray}
\begin{document}

\title{Transition from ordered pinched to warped magnetic field on a 100 au scale in the Class 0 protostar B335}

\author{Hsi-Wei Yen}
\affiliation{Academia Sinica Institute of Astronomy and Astrophysics, 11F of Astro-Math Bldg, 1, Sec. 4, Roosevelt Rd, Taipei 10617, Taiwan}

\author{Bo Zhao}
\affiliation{Max-Planck-Institut f\"ur extraterrestrische Physik (MPE), Garching, Germany, 85748}

\author{Patrick Koch}
\affiliation{Academia Sinica Institute of Astronomy and Astrophysics, 11F of Astro-Math Bldg, 1, Sec. 4, Roosevelt Rd, Taipei 10617, Taiwan}

\author{Ruben Krasnopolsky}
\affiliation{Academia Sinica Institute of Astronomy and Astrophysics, 11F of Astro-Math Bldg, 1, Sec. 4, Roosevelt Rd, Taipei 10617, Taiwan}

\author{Zhi-Yun Li}
\affiliation{Astronomy Department, University of Virginia, Charlottesville, VA 22904, USA}

\author{Nagayoshi Ohashi}
\affiliation{Academia Sinica Institute of Astronomy and Astrophysics, 11F of Astro-Math Bldg, 1, Sec. 4, Roosevelt Rd, Taipei 10617, Taiwan}

\author{Hsien Shang}
\affiliation{Academia Sinica Institute of Astronomy and Astrophysics, 11F of Astro-Math Bldg, 1, Sec. 4, Roosevelt Rd, Taipei 10617, Taiwan}

\author{Shigehisa Takakuwa}
\affiliation{Department of Physics and Astronomy, Graduate School of Science and Engineering, Kagoshima University, 1-21-35 Korimoto, Kagoshima, Kagoshima 890-0065, Japan}
\affiliation{Academia Sinica Institute of Astronomy and Astrophysics, 11F of Astro-Math Bldg, 1, Sec. 4, Roosevelt Rd, Taipei 10617, Taiwan}

\author{Ya-Wen Tang}
\affiliation{Academia Sinica Institute of Astronomy and Astrophysics, 11F of Astro-Math Bldg, 1, Sec. 4, Roosevelt Rd, Taipei 10617, Taiwan}

\correspondingauthor{Hsi-Wei Yen}
\email{hwyen@asiaa.sinica.edu.tw}

\begin{abstract}
We present our observational results of the 0.87 mm polarized dust emission in the Class 0 protostar B335 obtained with the Atacama Large Millimeter/submillimeter Array (ALMA) at a 0\farcs2 (20 au) resolution. We compared our data at 0.87 mm with those at 1.3 mm from the ALMA archive. The observed polarization orientations at the two wavelengths are consistent within the uncertainty, and the polarization percentages are systematically higher at 1.3 mm than 0.87 mm by a factor of $\sim$1.7, suggesting that the polarized emission originates from magnetically aligned dust grains. We inferred the magnetic field orientations from the observed polarization orientations. We found that the magnetic field changes from ordered and highly pinched to more complicated and asymmetric structures within the inner 100 au scale of B335, and the magnetic field connects to the center along the equatorial plane as well as along the directions which are $\sim$40$\arcdeg$--60$\arcdeg$ from the equatorial plane. We performed non-ideal MHD simulations of collapsing dense cores. We found that similar magnetic field structures appear in our simulations of dense cores with the magnetic field and rotational axis slightly misaligned by 15$\arcdeg$ but not in those with the aligned magnetic field and rotational axis. Our results suggest that the midplane of the inner envelope within the inner 100 au scale of B335 could be warped because of the misaligned magnetic field and rotational axis, and the magnetic field could be dragged by the warped accretion flows. 

\end{abstract}


\section{Introduction}
Knowing the relative importance of gravitational, rotational, turbulent, and magnetic energies is essential to understand the process of star formation,
as all of these effects shape structures and affect disk formation in protostellar sources \citep{McKee07, Li14a}.
Theoretical studies show that if the magnetic field and rotational axis of a dense core are aligned, 
magnetic braking is very efficient during the collapse in the ideal magnetohydrodynamic (MHD) condition \citep{Galli06, Allen03, Mellon08}.
As a consequence, the disk formation and growth are suppressed, 
and Keplerian disks with sizes larger than 10 au unlikely form around protostars.
When other mechanisms are incorporated, such as the non-ideal MHD effects, misaligned magnetic field and rotational axes in dense cores, or turbulence, 
the efficiency of magnetic braking is reduced, 
and a large Keplerian disk with a size of tens of au can form around a protostar in numerical simulations \citep{Dapp12,Joos12,Joos13,Santos12,Seifried13,Li13,Krumholz13,Tsukamoto15,Tsukamoto17,Zhao16,Zhao18,Masson16,Matsumoto17,Wurster19a,Wurster19b}.
Keplerian disks with sizes of tens of au have been observed around several young protostars \citep{Murillo13,Lee14,Lee18,Sakai14,Yen17}.
However, it remains unclear as to which mechanisms play more important roles to regulate the disk formation and growth in protostellar sources. 

B335 is an isolated Bok globule with an embedded Class 0 protostar at a distance of 100 pc \citep{Keene80, Keene83, Stutz08, Olofsson09}.
B335 is slowly rotating \citep{Saito99,Yen11,Kurono13}, and is associated with a 0.1 pc bipolar molecular outflow as well as several Herbig Haro objects along the east--west direction \citep{Hir88,Gal07}. 
The infalling motion has been observed on scales from 3000 au to 10 au in B335 \citep{Zhou93, Zhou95, Choi95, Evans05, Evans15, Saito99, Yen10, Yen11, Yen15, Kurono13, Imai19, Bjerkeli19}.
On the other hand, the rotational motion in the protostellar envelope on scales from 1000 au to 10 au in B335 has been found to be slow \citep{Yen10, Yen15, Imai19, Bjerkeli19}, 
and no clear sign of Keplerian rotation is observed even on a 10 au scale \citep{Bjerkeli19}. 
The James Clerk Maxwell Telescope (JCMT) and Atacama Large Millimeter/submillimeter Array (ALMA) polarimetric observations show that the magnetic field is along the east--west direction, and it is ordered and highly pinched on a 1000 au scale \citep{Maury18,Yen19}. 
These results suggest that the magnetic field likely plays an important role in the dynamics in B335. 
On the other hand, the presence of the jets and outflows in B335 could hint at the existence of a circumstellar disk \citep[e.g.,][]{Blandford82}.
In addition, a compact continuum component with a size of $\sim$6 au oriented perpendicular to the outflow direction as well as velocity gradients due to the rotational motion on a similar scale have been observed with ALMA \citep{Imai19,Bjerkeli19}.
Thus, a small disk with a size of a few au likely has formed in B335. 
Therefore, B335 is an excellent target to study the disk formation under the influence of a dynamically important magnetic field. 

To observationally investigate the effects of the magnetic field on the disk formation in B335, 
we conducted ALMA polarimetric observations at an angular resolution of 0\farcs2 (20 au) to probe the magnetic field structures on a scale close to the disk forming region. 
In the present paper, we describe the details of our observations in Section \ref{observations}, 
and introduce our observational results in Section \ref{results}. 
In addition, we compare our polarization data at 0.87 mm with those at 1.3 mm retrieved from the ALMA archive. 
In Section \ref{discussion}, we present the results from our non-ideal MHD simulations for B335, 
and discuss the observed magnetic field structures in comparison with the simulations. 
Lastly, in Section \ref{summary} we summarize the possible effects of turbulence as well as the misalignment between the magnetic field and rotational axis on the magnetic field structures in B335.

\section{Observations}\label{observations}
The polarimetric observations with ALMA at 0.87 mm toward B335 were conducted during 2016 to 2018, consisting of 13 successful executions (project code: 2015.1.01018.S). 
In the observations, 40 to 47 antennae were used in the configurations with baseline lengths ranging from 15 m to 1400 m. 
The pointing center was $\alpha({\rm J2000}) = 19^{\rm h}37^{\rm d}00.\!\!^{\rm s}89$, $\delta({\rm J2000}) =  +7\arcdeg34\arcmin09\farcs6$.
The on-source integration time was 7.4 hours. 
The observations were conducted with the full polarization mode and at the frequency ranges of 335.5--339.5 GHz and 347.5--351.5 GHz with a total bandwidth of 8 GHz.
In these observations, J1751$-$0939 was observed as the bandpass calibrator, J1938$+$0448 or J1935$+$2021 as the gain calibrators, and 
J1924$-$2914 or J2000$-$1748 for polarization calibration. 
The flux calibration was performed with the observations of quasars or the asteroid, Pallas. 
The data were manually calibrated by the EA ARC node using the Common Astronomy Software Applications (CASA) of the version 5.1.1 \citep{McMullin07}. 
We additionally performed self-calibration of the phase using the Stokes {\it I} data.
Then the calibrated visibility data were Fourier-transformed with the Briggs weighting with a robust parameter of +0.5 to generate Stokes {\it IQU} images, and the images were cleaned using the CASA task {\it tclean}.
The achieved synthesized beam is $0\farcs19 \times 0\farcs17$.
The noise level in the Stokes {\it I} image is 40 $\mu$Jy beam$^{-1}$, 
and it is 9 $\mu$Jy beam$^{-1}$ in the Stokes {\it Q} and {\it U} images.
When we generated the polarized intensity ($I_{\rm p}$) map, we debiased the polarized intensity ($I_{\rm p}$) with $I_{\rm p} = \sqrt{Q^2 + U^2 - {\sigma_{Q,U}}^2}$, 
where $\sigma_{Q,U}$ is the noise level in Stokes {\it Q} and {\it U} \citep{Wardle74,Simmons85}.
To extract polarization detections, we first binned up the Stokes {\it IQU} and $I_{\rm p}$ maps to have a pixel size of 0\farcs1, 
which is approximately half of the beam size, and computed polarization orientations and fractions.
Thus, the minimal separation between two polarization detections is 0\farcs1. 
The Stokes {\it I} and $I_{\rm p}$ maps with their original pixel size of 0\farcs02 are presented below.
The polarization detections are extracted when the signal-to-noise ratios (S/N) of both Stokes {\it I} and $I_{\rm p}$ are larger than three, 
and thus the expected uncertainties in the polarization orientations are $\lesssim$9\arcdeg.

\section{Results}\label{results}
\begin{figure*}
\centering
\includegraphics[width=\textwidth]{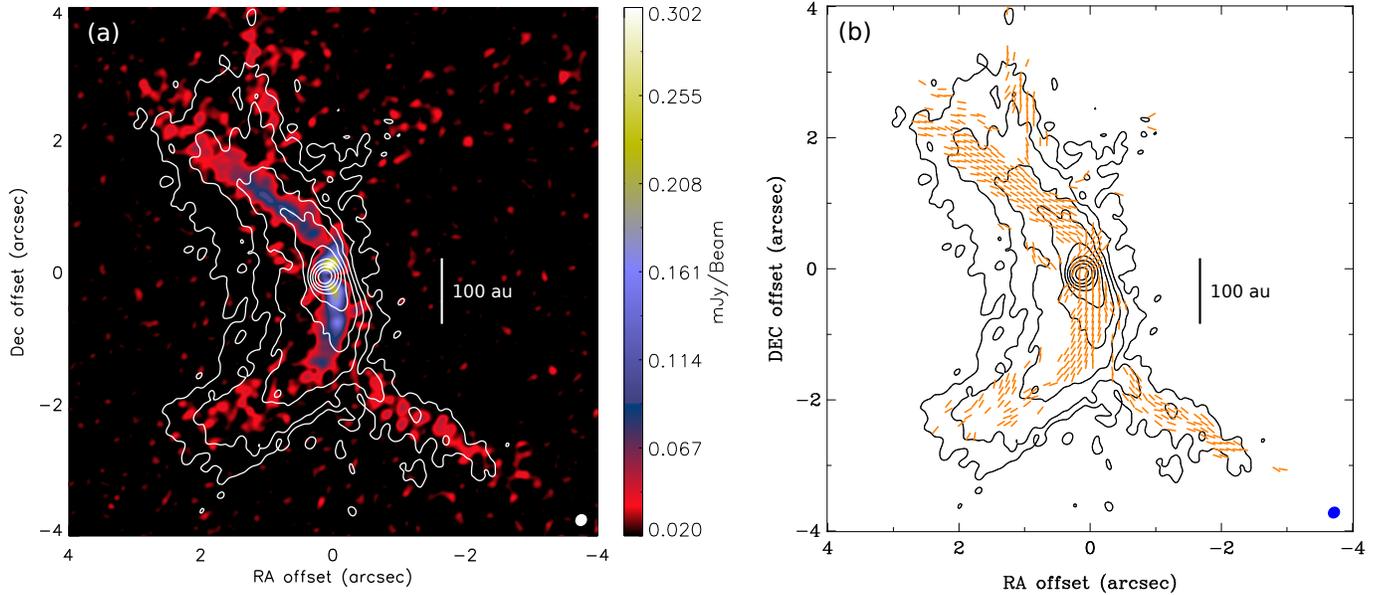}
\caption{(a) 0.87 mm continuum map (contours) overlaid on the polarized intensity map (color) obtained with our ALMA observations. The polarized intensity map is in units of mJy Beam$^{-1}$. The contour levels are from 5$\sigma$ in steps of a factor of two, where 1$\sigma$ is 40 $\mu$Jy Beam$^{-1}$.
(b) Magnetic field orientations (orange segments) inferred by rotating the polarization orientations by 90$\arcdeg$. The contours show the Stokes {\it I} intensity, identical to panel (a). The length of all the segments is the same. The minimal separation between two segments is half of the beam size. 
The zero offset refers to the pointing center ($19^{\rm h}37^{\rm d}00.\!\!^{\rm s}89 +7\arcdeg34\arcmin09\farcs6$). The filled ellipses at the bottom right corners present the size of the synthesized beam.}\label{polmap}
\end{figure*}

In the 0.87 mm continuum map obtained with our ALMA observations, 
there is a compact continuum source at the center with a peak position of $\alpha({\rm J2000}) = 19^{\rm h}37^{\rm d}00.\!\!^{\rm s}9$, $\delta({\rm J2000}) =  +7\arcdeg34\arcmin09\farcs5$ (Fig.~\ref{polmap}a). 
This peak position is consistent with the one measured at 1.3 mm at a higher angular resolution of 0\farcs03 \citep{Bjerkeli19}.
The apparent size of this compact continuum source in our observations is 0\farcs2 (20 au).
Thus, it is only marginally resolved, and we could not constrain its orientation. 
The peak brightness temperature is measured to be 30 K, 
which is more than a factor of two lower than the dust temperature of 67 K at a radius of 25 au estimated by modeling molecular-line and continuum data and the spectral energy distribution of B335 \citep{Shirley11,Evans15}.
Therefore, the continuum emission detected with our observations is most likely optically thin.

The compact source is embedded in a flattened structure with an apparent size of $\sim$2$\arcsec$ and a position angle (PA) of its major axis of 17\arcdeg. 
On a larger scale of $\sim$4$\arcsec$ (400 au), 
the continuum emission shows a bent structure, 
and its distribution is along the outflow cavity wall. 
In addition, 
the continuum emission on the 400 au scale in B335 lies primarily beyond the edge of the outflow traced by the CO (2--1) emission,
as seen in Fig.~5 in \citet{Bjerkeli19}.
The overall structures in the continuum emission in our observations are consistent with the previous results obtained with ALMA at 1.3 mm and angular resolutions of 0\farcs03 to 0\farcs8 \citep[e.g.,][]{Yen15,Imai16,Maury18,Bjerkeli19}.

The polarized continuum emission at 0.89 mm is primarily detected along the outflow cavity and around the central compact continuum source. 
The polarized intensity is highest in the south and north close to the Stokes {\it I} intensity peak but becomes lower at the peak (Fig.~\ref{polmap}a). 
As discussed in \citet{Maury18}, 
more polarized continuum emission detected along the outflow cavity wall could be due to more efficient dust grain alignment caused by stronger illumination from the central star in the outflow cavity \citep{Lazarian07}.
Similar enhancements in polarized continuum emission along outflow cavity walls have also been observed in other protostellar sources \citep[e.g.,][]{Hull17b,Hull19,Gouellec19}.
The orientations of the polarization detections are rotated by 90$\arcdeg$, displaying the orientations of the magnetic field (Fig.~\ref{polmap}b).
As discussed in Section \ref{discussion}, 
the detected polarized emission is unlikely induced by dust scattering based on the measured polarization fractions \citep{Kataoka15, Yang16, Yang17}.
The magnetic field orientations on a 400 au scale are primarily along the outflow cavity wall. 
There is a northern patch of detections with orientations more along the north--south direction. 
These features on a scale larger than 100 au are consistent with those observed at 1.3 mm with ALMA at an angular resolution of 0\farcs8 \citep{Maury18,Yen19}.

\begin{figure}
\centering
\includegraphics[width=0.5\textwidth]{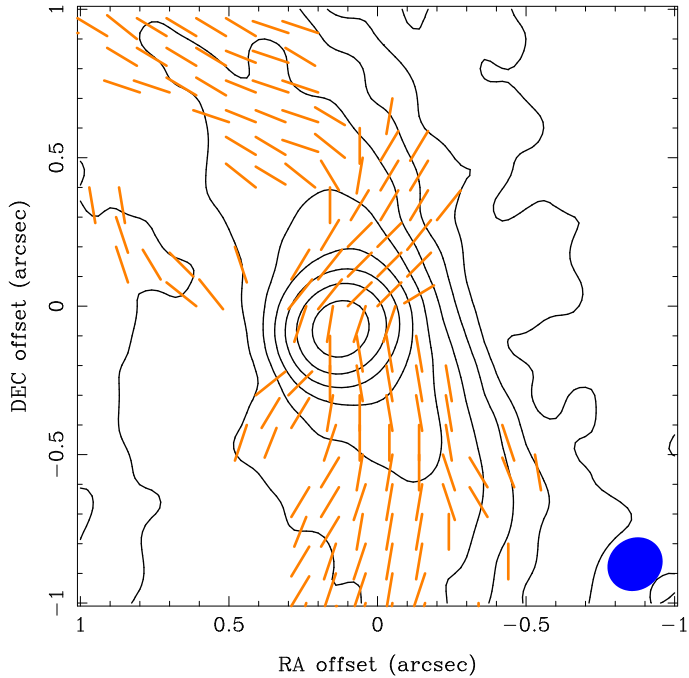}
\caption{Same as Fig.~\ref{polmap} (b) but zooming in on the central 2$\arcsec$~$\times$~2$\arcsec$ region.}\label{bmap}
\end{figure}

With a three times higher resolution, 
our observations reveal the magnetic field structures significantly closer to the central source than was possible with the previous, lower-resolution observations \citep{Maury18,Yen19}. 
In the present paper, we focus on the features on a small scale within radii of 0\farcs5--1$\arcsec$ (50--100 au). 
The magnetic field structures on a larger scale in B335 have been discussed in detail in \citet{Maury18} and \citet{Yen19}. 
Figure \ref{bmap} presents the magnetic field orientations in the central 2$\arcsec$ (200 au) region. 
At offsets of $\sim$0\farcs5--1$\arcsec$ in the north, 
the detected magnetic field orientations are along the wall of the outflow cavity. 
In the inner northern region at offsets of $\sim$0--0\farcs5, 
the magnetic field is along the northwest--southeast direction with PA of 130\arcdeg--150\arcdeg.
Here, the magnetic field segments appear to connect to the continuum peak, 
with a few segments also along the north--south direction with PA of 0\arcdeg--10\arcdeg.
In the south, the magnetic field is primarily along the north--south direction, 
and the magnetic field becomes more tilted from the north--south direction by 20\arcdeg--30\arcdeg at outer radii of $\sim$0\farcs5--1\arcsec. 
In addition, there are a few magnetic field segments along the northwest--southeast direction detected at $\sim$0\farcs4 southeast from the continuum peak.
In the east, at RA offsets of 0\farcs6--1$\arcsec$, there is a patch of segments with their PA gradually changing from 10$\arcdeg$ to 60$\arcdeg$ toward the center.

\begin{figure*}
\centering
\includegraphics[width=\textwidth]{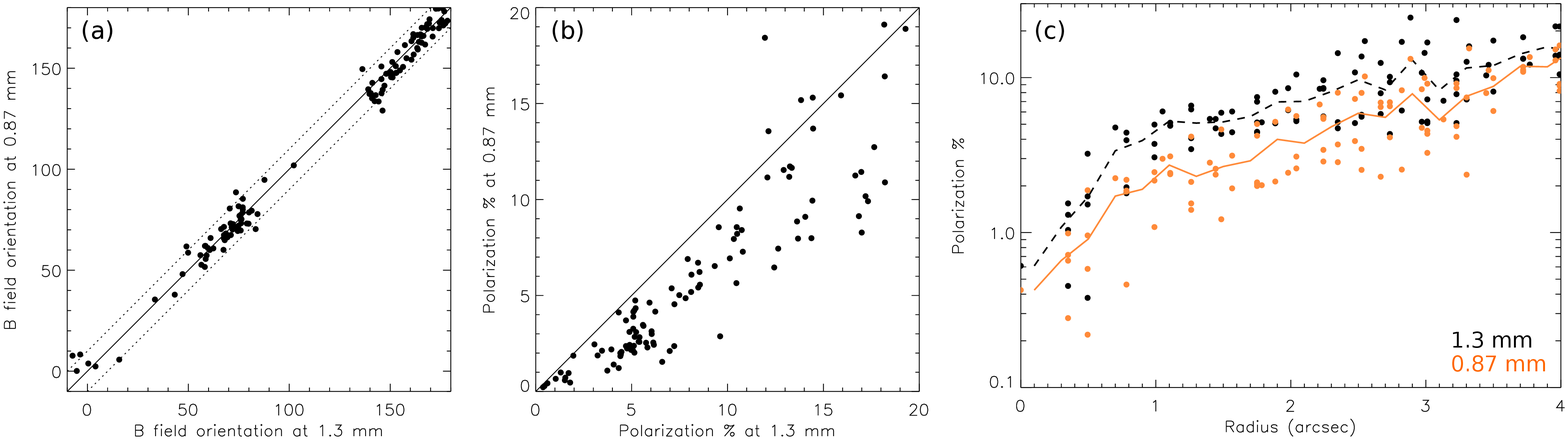}
\caption{Comparison of (a) the magnetic field orientations and (b) the polarization percentages inferred from the polarization data at 0.87 mm and 1.3 mm. The data were convolved to the same beam size. The dotted lines in the panel (a) show the difference of $\pm$10\arcdeg. (c) presents the polarization percentages at 0.87 mm (orange) and 1.3 mm (black) as a function of projected radius with respect to the continuum peak. Orange solid and black dashed curves show the running means in radial bins of 0\farcs2.}\label{2pol}
\end{figure*}

For comparison, we retrieved the polarization maps at 1.3 mm obtained with ALMA \citep{Maury18,Yen19}. 
We convolved our polarization maps at 0.87 mm to have the same beam size as those at 1.3 mm and extracted polarization detections with the same procedure and criteria as those in \citet{Yen19}. 
Then we compared the polarization orientations and percentages at 0.87 mm and 1.3 mm at the same positions. 
Figure \ref{2pol} presents the comparison of the inferred magnetic field orientations and measured polarization percentages at 0.87 mm and 1.3 mm.
The magnetic field orientations inferred from the polarization data at 0.87 mm and 1.3 mm are consistent within the uncertainties. 
The mean difference in the orientations at the two wavelengths is 5$\arcdeg$ with a standard deviation of 4$\arcdeg$. 
More than 90\% of the polarization detections show higher polarization percentages at 1.3 mm than at 0.87 mm. 
On average, the polarization percentages at 1.3 mm are a factor of 1.7 higher.
We note that the polarization percentages at both wavelengths increase with increasing radii, 
and the percentages at 0.87 mm are systematically lower than those at 1.3 mm at all radii (Fig.~\ref{2pol}c). 

\section{Discussion}\label{discussion}

\subsection{Origins of the polarized continuum emission}
With the ALMA polarization data at 0.87 mm and 1.3 mm, 
we found higher polarization percentages at the longer wavelength.
This is the opposite of the expected trend for the scattering-induced polarization by dust grains with sizes up to $\sim$100 $\mu$m, 
where the scattering cross-section drops steeply with wavelength as $\lambda^{-4}$ \citep{Kataoka15, Yang16}, 
unless the region is optically thick. 
In addition, the observed polarization fraction is higher than typically predicted by scattering models. 
In B335, the continuum emission on the scale which we observe is optically thin (Section \ref{results}). 
Thus, the polarized continuum emission down to a 20 au scale in B335 most likely originates from magnetically aligned grains. 
Therefore, the magnetic field structures in B335 can be inferred by rotating the observed polarization orientations by 90\arcdeg. 

The polarization percentages as a function of wavelength could depend on the compositions and sizes of dust grains \citep[e.g.,][]{Bethell07,Draine09,Ashton18,Valdivia19}.
Among a few dust models that reproduce the spectral energy distributions and extinction curve of the interstellar medium (ISM), 
the one with spheroidal silicate and spherical carbon grains shows increasing polarization percentages from submillimeter to millimeter wavelengths \citep{Draine09}, 
similar to the trend observed in B335.
Nevertheless, the properties of dust grains and the environments, such as grain sizes and radiation fields, are expected to be different in the ISM and dense cores \citep{Bethell07,Ashton18}.
Consequently, ISM dust models might not be directly applicable to B335. 

More specifically, 
\citet{Valdivia19} computed polarized radiative transfer in typical protostellar envelopes with the radiative torque theory \citep{Lazarian07} and dust size distributions similar to the Mathis-Rumpl-Nordsieck (MRN) distribution \citep{Mathis77}.
They further studied the dependences of the polarization percentages at 0.8 mm and 1.3 mm on the maximum grain sizes.
They found that the presence of dust grains with sizes larger than 10 $\mu$m in protostellar envelopes on a scale of a few hundred au is needed to explain the observed polarized percentages of a few to more than 10\% in protostellar sources.
In addition, in their model calculations with maximum grain sizes of 30--50 $\mu$m in protostellar envelopes, 
the polarization percentages on a scale of a few hundred au are higher at 1.3 mm than at 0.8 mm. 
These trends are consistent with our observational results of B335.
Thus, our results may hint at the presence of large dust grains with sizes of a few tens of $\mu$m on a 100 au scale in B335.

\subsection{Magnetic field structures}
The previous JCMT and ALMA polarimetric observations \citep{Maury18, Yen19} show that the magnetic field structures in B335 are ordered and along the outflow (east--west) direction on the scale of the natal dense core of 6000 au, and then become highly pinched on the scale of the infalling envelope of 1000 au. 
In addition, close to the center, the magnetic field orientations 
are almost along the north--south direction, which is the direction of the disk midplane. 
These observed magnetic field structures can be explained with non-ideal MHD simulations of a collapsing dense core with a weak magnetic field of mass-to-flux ratios of $\sim$6--10 aligned with the rotational axis \citep{Maury18, Yen19}.

With our observations at higher angular resolutions, 
the magnetic field along the north--south direction is detected within a 100 au scale north and south to the center. 
In addition to these field lines along the direction of the disk midplane, 
there are also magnetic field lines tilted with respect to the midplane in the 100 au region around the central compact source (Fig.~\ref{bmap}).
Our observations also show that the orientation of its surrounding flattened envelope on a 100 au scale is not aligned with the disk midplane but is tilted eastwards by $\sim$20$\arcdeg$ (Fig.~\ref{bmap}).  
Such more complicated magnetic field structures around the central disk and the different orientations between the flattened envelope and the disk are not expected in the non-turbulent simulations of a collapsing dense core with a magnetic field and rotational axis that are aligned \citep[e.g.,][]{Li14b, Masson16, Tsukamoto17, Zhao18, Machida19}.

\subsubsection{Simulation with the aligned magnetic field and rotational axis}
\begin{figure*}
\centering
\includegraphics[width=\textwidth]{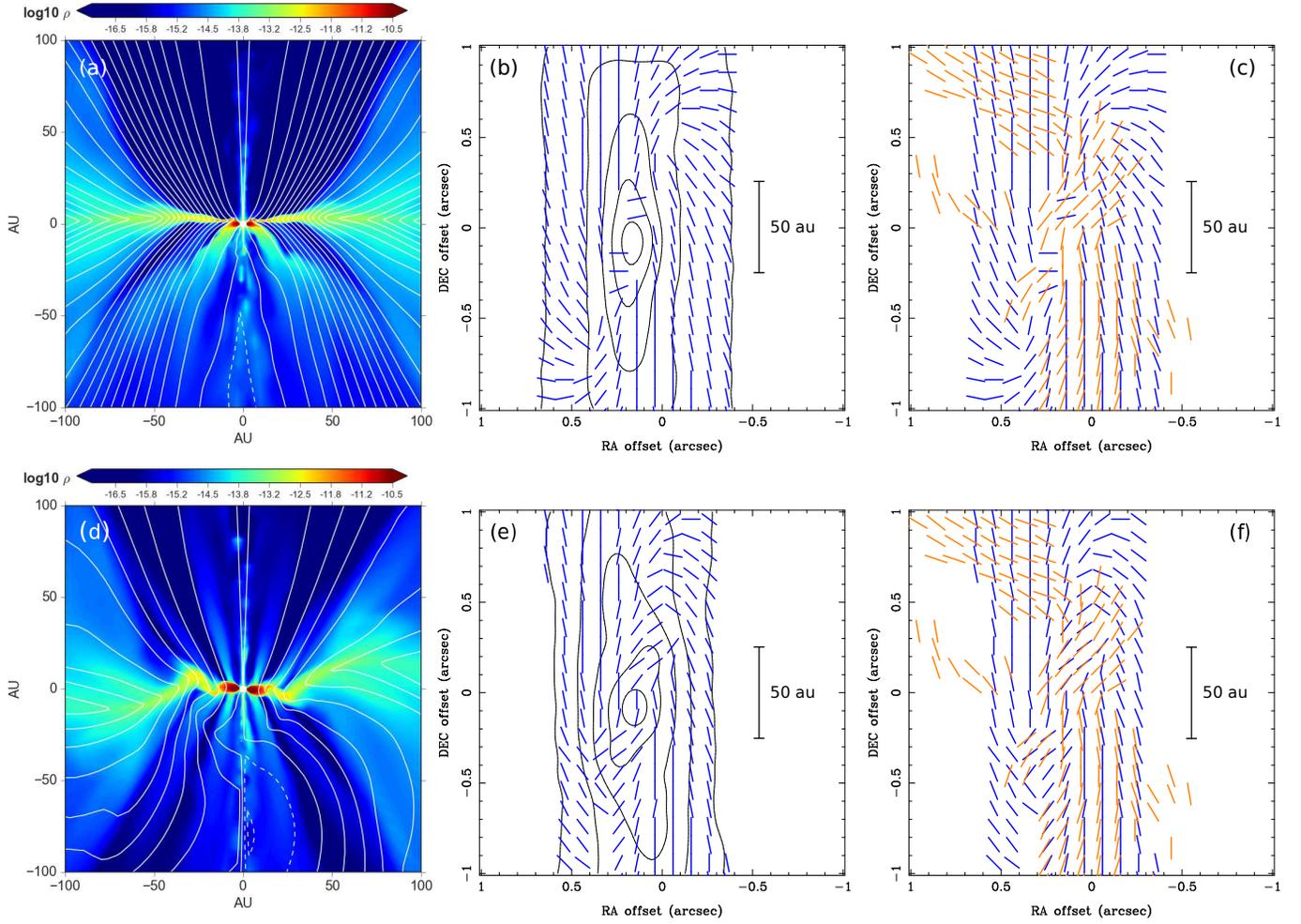}
\caption{(a) Density (color; in units of g~cm$^{-3}$) and magnetic field lines (white lines) in our non-ideal MHD simulation of a collapsing dense core with the aligned magnetic field and rotational axis. The outflow direction is along the vertical direction. (b) Synthetic ALMA observational results of our simulation in (a). The synthetic observations are rotated to have an inclination angle of 80$\arcdeg$ and an outflow direction along the east--west direction, the same as in B335. The contours show the distribution of the Stokes {\it I} intensity, and the segments show the magnetic field orientations inferred by rotating the polarization orientations by 90$\arcdeg$. (c) Comparison of the observed (orange) and model (blue) magnetic field orientations. The blue segments are the same as those in (b). Panels (d)--(f) are the same as (a)--(c) but for our simulation of a collapsing dense core with its magnetic field misaligned with the rotational axis by 15\arcdeg. The outflow direction is also along the vertical direction in (d). In (a) and (d), the magnetic field lines were plotted using iso-magnetic flux contours as approximation, and the dashed contours near the polar axis denote the regions with negative magnetic fluxes, where the structures of the magnetic field lines cannot be properly evaluated in two-dimensional plots.}\label{simob}
\end{figure*}

We have compared the magnetic field structures detected with our ALMA observations with those in the non-ideal MHD simulation of a collapsing dense core with the aligned magnetic field and rotational axis. 
The synthetic {\it IQU} maps of the simulation results were obtained from \citet{Yen19}. 
Among the three non-ideal MHD effects, the simulation only includes ambipolar diffusion, which is the most dominant non-ideal MHD effect in protostellar envelopes on a scale of hundreds of au \citep{Zhao18}.
In this simulation, the dense core has the same mass and angular velocity as the observational estimates of B335, and an initial mass-to-flux ratio of 9.6. 
A cosmic ray ionization rate of $5 \times 10^{-17}$ was adopted.
The simulation was stopped when the central stellar mass reached 0.1 $M_\sun$. 
At the end of our simulation, the disk size was $<$10 au. 
These stellar mass and disk size are comparable to the observational estimates of B335 \citep{Evans15, Yen15, Bjerkeli19, Imai19}.
The synthetic {\it IQU} maps were generated using the radiative transfer code, Simulation Package for Astronomical Radiative Xfer (SPARX; \url{https://sparx.tiara.sinica.edu.tw/}), on the assumption of a constant polarization efficiency meaning that the polarized intensity is simply proportional to the column density.   
The details of the simulation are described in \citet{Yen19}. 
This simulation can explain observed magnetic field structures on a few hundred au to 6000 au scales in B335. 
We then synthetically observed the Stokes {\it IQU} maps output by SPARX using the CASA simulator. 
The resulting simulated maps are presented in Fig.~\ref{simob} (a)--(c). 

In the simulation with the aligned magnetic field and rotational axis, 
the Stokes {\it I} intensity from large to small scale is all elongated along the direction of the disk midplane, which is the north--south direction (Fig.~\ref{simob}b). 
The magnetic field inferred from the synthetic polarization maps is perpendicular to the disk midplane within 0\farcs2 in the north and south of the intensity peak. 
Then, it becomes almost parallel to the disk midplane at outer radii in the north and the south (Fig.~\ref{simob}b). 
Thus, this simulation can explain the magnetic field orientations parallel to the disk midplane observed in the north and the south on a 100 au scale in B335. 
Especially, it reproduces most of the observed magnetic field orientations in the south (Fig.~\ref{simob}c). 
However, the magnetic field near the intensity peak in the synthetic maps is not along the northwest--southeast direction,
which is different from the observations.  
Thus, the observed magnetic field structures within a 100 au scale cannot be fully explained with the simulation of a collapsing dense core with the aligned magnetic field and rotational axis, although this simulation was able to explain the pinched magnetic field observed on scales from 6000 au down to a few hundred au \citep[Fig.~\ref{jcmt_simob} upper; ][]{Yen19}.

\subsubsection{Simulation with the misaligned magnetic field and rotational axis}
On the contrary, in simulations of a collapsing dense core with the misaligned magnetic field and rotational axis or with turbulence, 
the magnetic field structures tend to become irregular, different from a simple hourglass shape \citep[e.g.,][]{Li14b, Seifried15, Masson16, Hull17a, Tsukamoto17, Wurster19a}.
Thus, we additionally performed non-ideal MHD simulations of a collapsing dense core with its magnetic field misaligned with its rotational axis by 15\arcdeg.
The mass, angular velocity, and initial mass-to-flux ratio of the dense core are the same as those in our simulation with the aligned magnetic field and rotational axis.  
Due to the slight misalignment, the disk is easier to grow in size compared to the simulation without the misalignment. 
Thus, in this simulation with the misalignment, 
the cosmic-ray ionization rate was increased to be $1\times10^{-16}$~s$^{-1}$, which enhances the coupling between the magnetic field and the matter, 
and the disk size was also suppressed to be $<$10 au to match the observational constraint of the disk size in B335.
The simulation was stopped at a similar evolutionary time when the central stellar mass reached 0.1 $M_\sun$.
Thus, the only differences between the initial conditions of these two simulations are the cosmic-ray ionization rates and the angles between the magnetic field and rotational axis.
Finally, we produced synthetic observations using the CASA simulator in the same manner as for the simulations with the aligned magnetic field and rotation axis.

The synthetic maps from our simulation with the misaligned magnetic field and rotational axis are presented in Fig.~\ref{simob}(d)--(f). 
Because the rotational axis and magnetic field are misaligned, 
the flattened envelope becomes warped on the small scale of 100 au (Fig.~\ref{simob}d), 
where the rotational energy becomes also important compared to the gravitational and magnetic energy. 
As a result, the elongations of the Stokes {\it I} intensity distributions in the synthetic map change from large to small scales (Fig.~\ref{simob}e). 
The magnetic field is dragged by the accretion flows along the warped envelope, resulting in more complicated magnetic field structures. 
We note that this simulation with the misalignment has a higher cosmic ray ionization rate.
In the simulations, the magnetic field structures are shaped by the gas motions.
Increasing the cosmic ray ionization rate does not directly affect the magnetic field structures but only strengthens the coupling between the gas motions and the magnetic field \citep{Yen19}. 
Thus, the complicated magnetic field structures seen in this simulation is due to the changes in the gas motion caused by the misalignment.

In the synthetic map, the magnetic field is parallel to the disk plane at the intensity peak and is along the northwest--southeast direction within 0\farcs5 around the intensity peak. 
Then, the magnetic field becomes more aligned along the north--south direction in the outer regions (Fig.~\ref{simob}e). 
These features are similar to what is observed on a 100 au scale in B335,
although the regions showing the magnetic field along the northwest--southeast direction extend further away from the disk midplane in the observations compared to our simulation. 
Outside the central 100 au region, the magnetic field structures in the synthetic maps of the simulations with and without misalignment are similar (Fig.~\ref{simob}b \& e).
Thus, our new simulation with the misalignment can also explain the magnetic field structures in the protostellar envelope on a scale of a few hundred au in B335 detected with the previous ALMA observations at a lower spatial resolution of 70--80 au \citep{Maury18,Yen19}.

\subsubsection{Comparison between the aligned and misaligned cases}
\begin{figure}
\centering
\includegraphics[width=0.45\textwidth]{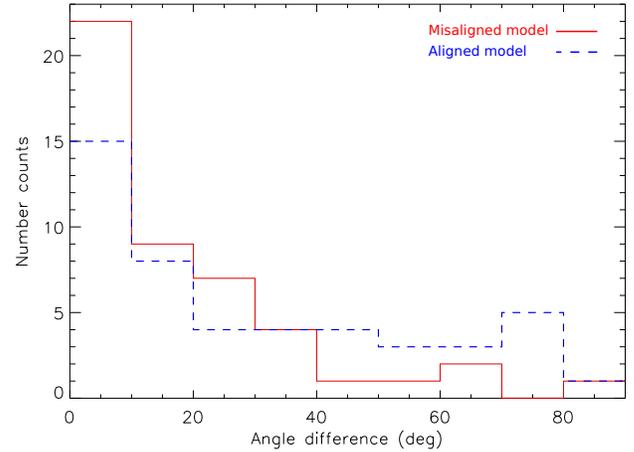}
\caption{Number distributions of angle difference between observed and model magnetic field orientations in the central 100 au region. Red solid and blue dashed histograms show the cases for the simulations with the magnetic field misaligned and aligned with the rotational axis, respectively.}\label{diff}
\end{figure}

In Figure \ref{diff}, we present the number distributions of the angle difference of the magnetic field orientations in the central 100 au region extracted from the observed and synthetic polarization data. 
The number of the model segments with their orientations consistent with the observations within 30$\arcdeg$ increases by 50\%, 
when we include the misalignment in our simulations.
Thus, the simulation with the misalignment can explain the observed magnetic field orientations better than the simulation without the misalignment.

\begin{figure}
\centering
\includegraphics[width=0.4\textwidth]{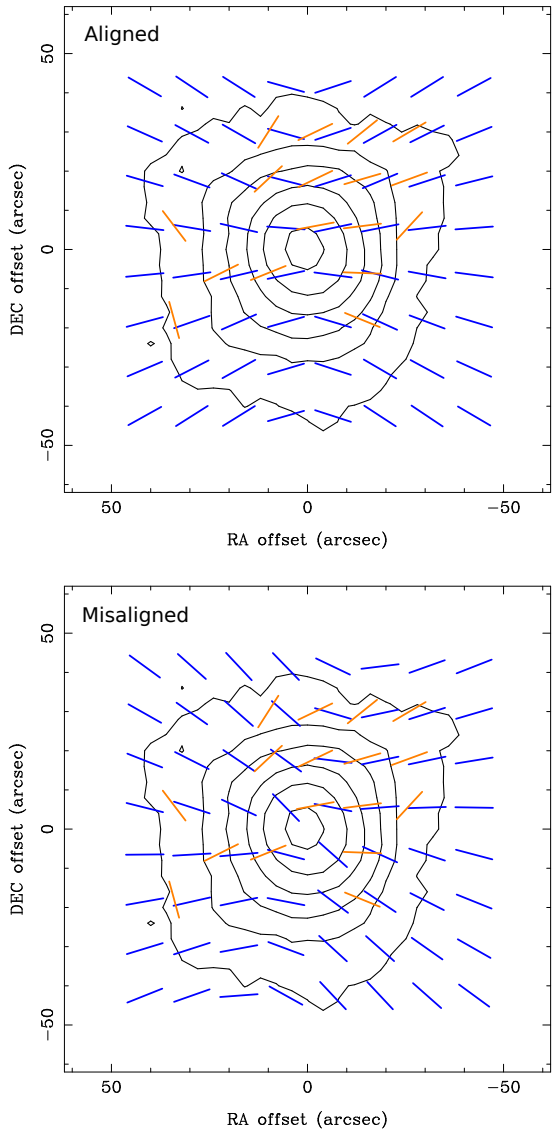}
\caption{Comparison of the magnetic field orientations observed with JCMT (orange segments) and those from our non-ideal MHD simulations (blue segments) of collapsing dense cores with the magnetic field aligned (upper panel) and misaligned (lower panel) with the rotational axes. The contours show the Stokes {\it I} map of B335 obtained with JCMT. The JCMT polarimetric data were retrieved from \citet{Yen19} and were reduced again with the updated software and procedure. The S/N were improved with the new data reduction, and we obtained more detections than \citet{Yen19}.}\label{jcmt_simob}
\end{figure}

We have also compared the magnetic field structures on the large scale in this simulation with those observed with JCMT \citep{Yen19}. 
The degree of the misalignment between the magnetic field and rotational axis in the initial core in our simulation is only 15$\arcdeg$ in the three-dimensional space. 
Thus, after projection on the plane of the sky, the simulation with the misalignment (Fig.~\ref{jcmt_simob} lower panel) can reproduce the observed magnetic field orientations on the 6000 au scale in B335 similar to the simulation with the aligned magnetic field and rotation axis (Fig.~\ref{jcmt_simob} upper panel).
Hence, the critical difference between the aligned and the misaligned cases only manifests itself in the inner 100 au region.

We note that the intensity distribution is elongated along the northeast--southwest direction on a 1$\arcsec$ scale and then becomes along the northwest--southeast direction on a 0\farcs5 scale in our synthetic Stokes {\it I} map of the simulation with the misalignment (Fig.~\ref{simob}e).
This change in the intensity distribution is not fully consistent with the observations. 
The observed intensity distribution is elongated along the northeast--southwest direction on a 1$\arcsec$ scale, 
and then transitions into a roundish morphology on a 0\farcs5 scale. 
Our simulations were performed with the barotropic approximation, 
which could underestimate the temperature in the inner region of tens of au compared to radiative MHD simulations \citep{Tomida10}.
Consequently, the material could be more concentrated in the midplane in our simulations than in reality. 
Because the Stokes {\it I} intensity distribution is sensitive to the density distribution, 
this effect could make the warped structures more evident in our synthetic intensity map, compared to the observations.
On the contrary, the orientations of the magnetic field lines are more sensitive to the directions of the gas flows. 

In addition,
a constant polarization efficiency was adopted in our radiative transfer calculations, 
so our synthetic polarization maps primarily trace the magnetic field in the envelope. 
Thus, there are no magnetic field orientations along the outflow cavity wall detected in our synthetic maps, 
while such detections are seen in the northeast in the observed maps (Fig.~\ref{bmap}). 
We also note that the magnetic field structures and density distributions in our simulations (even for the misaligned case) are more or less axisymmetric. 
Thus, the toroidal components of the magnetic field along the line of slight may be canceled out in our radiative transfer calculations, 
causing our synthetic polarization maps to show axisymmetric patterns (Fig.~\ref{simob} and \ref{jcmt_simob}).
However, the observations show the magnetic field orientations are different in the north and the south of the center in B335, 
which is not seen in our synthetic maps and could suggest the presence of asymmetric structures in the magnetic field or the density distributions along the line of slight. 
This observed asymmetry is often seen in numerical simulations with turbulence \citep[e.g.,][]{Li14b, Masson16, Matsumoto17, Wurster19a}. 
Thus, in addition to seeing magnetic field perturbations caused by the misalignment of the magnetic field and the rotation axis, 
we may also be seeing the effect of gravo-turbulence in B335, 
which could cause asymmetries that become more evident on small scales as the core collapses.

Besides B335, 
ordered magnetic fields in a pinched hourglass shape have been observed on a 1000 au scale in several protostellar sources \citep{Girart06, Stephens13, Rao14, Cox18, Sadavoy18a, Sadavoy19, Kwon19, Ko19}. 
These results hint at an important role of the magnetic field in the dynamics in these protostellar sources. 
On the other hand, there are also protostellar sources showing disordered magnetic field structures on a 1000 au scale, 
which can be explained better with simulations with turbulence \citep{Hull17a, Hull17b, Cox18, Sadavoy18b, Sadavoy19}.
Our observations together with the previous JCMT and ALMA observations show that the magnetic field from large to small scales changes from the ordered pinched structures to more complicated and asymmetric structures in B335. 
A similar trend has also been seen in other protostellar sources, such as IRAS 16293$-$2422 \citep{Rao14,Sadavoy18b}.
These results suggest that the relative importance among magnetic field, rotation, and turbulence changes as a function of scale. 
Furthermore, it seems that the influence of rotation and turbulence, which are passed down from larger scales and enhanced by gravitational collapse, may become more significant in the inner envelopes around protostellar disks.

\section{Summary}\label{summary}
We present our observational results of the polarized 0.87 mm continuum emission at an 0\farcs2 resolution in B335 obtained with ALMA. 
We compare our results with those from the ALMA polarimetric observations at 1.3 mm with an 0\farcs8 resolution as well as with the synthetic maps generated from our non-ideal MHD simulations of collapsing dense cores. 
The main results are summarized below. 
\begin{enumerate} 
\item{
The polarization orientations at 0.87 mm and 1.3 mm are consistent within the uncertainties. 
The polarization percentage is higher at 1.3 mm than 0.87 mm on scales from 1000 au down to tens of au in B335, 
suggesting that the polarized emission originates from magnetically aligned dust grains.
In addition, the peak brightness temperature of the 0.87 mm continuum emission on a 20 au scale is 30 K, 
more than a factor of two lower than the expected dust temperature.
Thus, the 0.87 mm continuum emission is most likely optically thin down to a 20 au scale in B335, 
and the observed polarization orientations can be rotated by 90$\arcdeg$ to infer the magnetic field orientations.}
\item{Our observations show that the magnetic field structures in B335 change from an ordered pinched morphology on a 1000 au scale to more complicated structures on a 100 au scale. 
Within a 100 au scale, 
in addition to the magnetic field along the equatorial plane in the north and the south to the center, 
there are also magnetic field lines along the northwest--southeast direction that are connected to the central disk-forming region.}
\item{We have performed non-ideal MHD simulations of collapsing dense cores with their magnetic field aligned and misaligned with their rotational axes, 
and generated synthetic polarization maps to compare with the observations.
The simulation with the aligned magnetic field and rotational axis can explain the observed magnetic field orientations along the equatorial plane on a 100 au scale, but cannot explain those along the northwest--southeast direction connecting the central disk-forming region. 
On the contrary, 
the simulation with the misaligned magnetic field and rotational axis can reproduce both the magnetic field orientations along the equatorial plane and the northwest--southeast direction observed on a 100 au scale in B355.
The misalignment is 15$\arcdeg$ in our simulation, which is relatively small, 
and thus, this simulation can also explain the observed magnetic field orientations on a 6000 au scale with JCMT.}
\item{Our results suggest that the magnetic field and rotational axis in B335 are likely slightly misaligned. 
In addition, the observed different magnetic field orientations in the north and south on a scale within tens of au, 
which is not seen in our simulation with the misaligned magnetic field and rotation axis, could hint at the contribution of gravo-turbulence in B335.
Therefore, our observational results suggest that the relative importance between magnetic field, rotation, and turbulence changes as a function of scale in protostellar sources. 
On the small scale around the disk-forming region in B335, 
the rotational energy and influence by turbulence become more significant, 
and thus, the protostellar envelope is warped. 
The magnetic field lines could be dragged by the accretion flows along the warped envelope, resulting in more complicated and asymmetric structures. 
}
\end{enumerate}

\begin{acknowledgements} 
We thank Alfonso Trejo-Cruz for his efforts on the manual data calibration. 
We thank I-Ta Hsieh and Sheng-Yuan Liu for their assistance and advice on our radiative transfer calculation using SPARX.
This paper makes use of the following ALMA data: ADS/JAO.ALMA\#2015.1.01018.S. ALMA is a partnership of ESO (representing its member states), NSF (USA) and NINS (Japan), together with NRC (Canada), MOST and ASIAA (Taiwan), and KASI (Republic of Korea), in cooperation with the Republic of Chile. The Joint ALMA Observatory is operated by ESO, AUI/NRAO and NAOJ. 
We thank all the ALMA staff supporting this work. 
H.-W.Y. acknowledges support from MOST 108-2112-M-001-003-MY2.
P.M.K. acknowledges support from MOST 108-2112-M-001-012 and MOST 107-2119-M-001-023, and from an Academia Sinica Career Development Award.
ZYL is supported in part by NASA 80NSSC18K1095 and NSF AST-1716259 and 1815784.
S.T. acknowledges a grant from JSPS KAKENHI grant No. JP18K03703 in support of this work.
This work was supported by NAOJ ALMA Scientific Research grant No. 2017-04A.
\end{acknowledgements}

\software{CASA \citep[v5.1.1;][]{McMullin07}, SPARX (\url{https://sparx.tiara.sinica.edu.tw/)}}
\end{document}